\documentclass[11pt,twoside,a4paper,cr,final]{cms-tdr}
\def\svnVersion{144636M}

\begin{document}

\hyphenation{had-ron-i-za-tion}
\hyphenation{cal-or-i-me-ter}
\hyphenation{de-vices}

\RCS$Revision: 138024 $
\RCS$HeadURL: svn+ssh://svn.cern.ch/reps/tdr2/papers/XXX-08-000/trunk/XXX-08-000.tex $
\RCS$Id: XXX-08-000.tex 138024 2012-07-19 04:04:00Z alverson $
\newlength\cmsFigWidth
\ifthenelse{\boolean{cms@external}}{\setlength\cmsFigWidth{0.85\columnwidth}}{\setlength\cmsFigWidth{0.4\textwidth}}
\ifthenelse{\boolean{cms@external}}{\providecommand{\cmsLeft}{top}}{\providecommand{\cmsLeft}{left}}
\ifthenelse{\boolean{cms@external}}{\providecommand{\cmsRight}{bottom}}{\providecommand{\cmsRight}{right}}
\cmsNoteHeader{2012/174} % This is over-written in the CMS environment: useful as preprint no. for export versions
\title{W, Z and photon production in CMS}

\author*[Ciemat]{Bego\~{n}a de la Cruz on behalf of CMS Collaboration}

\date{\today}

\abstract{
   The production of electroweak bosons (photons, W and Z particles) in PbPb and pp collisions at sqrt(s) = 2.76 TeV per interacting nucleon pair has been measured with the CMS detector at the LHC. Direct photon production is studied using samples of isolated photons. W and Z bosons are reconstructed through their leptonic decay into muons. Their production rate in PbPb data is studied as a function of the centrality of the collision and compared to that in pp interactions, once normalized by the number of binary nucleon-nucleon interactions. The results are also compared to next-to-leading-order perturbative QCD calculations. 

\bigskip

Presented at {\em Hard Probes 2012: The 5th International Conference on Hard and Electromagnetic Probes of High-Energy Nuclear Collisions}
}

\hypersetup{%
pdfauthor={CMS Collaboration},%
pdftitle={CMS Paper Template 2006 LaTeX/PdfLaTeX version},%
pdfsubject={CMS},%
pdfkeywords={CMS, physics, software, computing}}

\maketitle %maketitle comes after all the front information has been supplied
\section{Introduction}

The measurement of electroweak bosons (photon, W and Z) represents one of the basic tests of the Standard Model of elementary particles. In the context of heavy ion collisions these bosons are not expected to interact with the hot and dense strongly interacting medium created. In that case, their production rate and properties can be taken as a reference for other particles which result affected or provide normalisation for other physics processes.
Thus, the direct comparison of the production cross section of such probes in pp and in nuclear collisions allows one to estimate possible modifications of nuclear parton density functions (PDF) relative to a simple incoherent superposition of nucleon PDF.

Photons have already been studied in heavy ion collisions at several experiments, at the Relativistic Heavy Ion Collider (RHIC)~\cite{Adler:2005ig} and, more recently, also at the Large Hadron Collider (LHC)~\cite{Chatrchyan:2011ip,CMS_gamma_jet}, but it has been the energy and luminosity conditions from the LHC that has made the detection of W and Z bosons possible. Studying these bosons in their leptonic decay modes turns out to be very useful, as leptons suffer negligible energy loss in the medium. Results on weak bosons production in PbPb collisions have already been published by ATLAS~\cite{Atlas:2010px} and CMS~\cite{Chatrchyan:2011ua, W-paper}  collaborations.

Several challenges are faced when studying the electroweak bosons in PbPb collisions. In the case of photon production, the main proccesses generating direct prompt photons (quark-gluon Compton scattering and quark-antiquark annihilation) are difficult to disentangle from the contribution of gammas from parton fragmentation processes, mainly $\pi^0$ and $\eta$ decays, giving rise to a wide variety of produced photons.

In the case of weak (W and Z) boson production, the main challenge lies in their low production cross section, compared to other processes. Although the leptonic decay modes are useful in identifying the boson production, they suppose an additional reduction factor ($\le 10\%$) of the yields.

The CMS experiment~\cite{CMS_det}  is well suited to detect muons and photons. In particular, for the photon detection and measurement, the electromagnetic calorimeter (ECAL) possesses a fine-grained ($\Delta\eta \times \Delta\Phi = 0.0174 \times 0.0174$ in $|\eta^{\gamma}|< 1.44$) hermetic disposition, arranged in a quasi-projective geometry, based on lead tungstate crystals which provide very narrow and precise electromagnetic showers.

The silicon pixel and strips tracker allows tracking and transverse momentum particle measurement in the pseudorapidity range $|\eta| < 2.5$, with a good momentum resolution between 2-3\% for charged tracks of up to 100 GeV, thanks to the high magnetic field (3.8T) provided by the CMS superconducting solenoid.

Muons are measured in the range $|\eta^{\mu}| < 2.4$. Their triggering and identification takes place in the muon chambers system, embedded in the steel return yoke of the magnet, with an excellent performance, while the muon transverse momentum is measured in the silicon tracker.

The centrality of the collisions is determined from the energy deposited in the forward steel/quartz-fibre calorimeters.
The measurements presented here are based on PbPb and pp data samples, both taken at $\sqrt{s} = 2.76$ TeV per interacting nucleon pair, corresponding to an integrated luminosity of around 7 $\mu$b$^{-1}$ and 231 nb$^{-1}$, respectively.
Similar measurements have also been performed in larger samples of pp collisions at $\sqrt{s} = $7 and 8  TeV, although they are not reported here.

\section{Isolated Photons}
\label{photons}

Photons coming directly from the hard scattering of incident nucleons are usually highly energetic and produced isolated from other particles. This is not the case for photons from background processes, mainly coming from decays of neutral mesons produced inside a jet, and thus, surrounded by significant hadronic activity from other parton fragments. Isolation provides a good handle to suppress the fragmentation photon component in the data sample, while removing a very small contribution of direct photons. Still, the hard scattering is superimposed on top of other multiple parton-parton scatterings (underlying event) occurring simultaneously and which must be subtracted, prior to applying the isolation criteria.

The contribution of the underlying event is subtracted, based on the mean value of the energy deposited per unit area along the $\eta-\phi$ strip containing the photon candidate in the ECAL. Once the mean energy is removed, the isolation requirements ($\sum E_{T} < 5$ GeV inside a cone defined by $\Delta$R= 0.4, considering the energy deposited in the electromagnetic and hadronic calorimeters and Tracker) is applied.
The resulting sample of isolated photons is dominated by quark-gluon Compton photons, with still a small proportion of photons from decays and fragmentation processes, which fulfill the isolation criteria.

The shape of the shower deposited in the ECAL, given by the transverse width of the cluster (second moment of the electromagnetic energy cluster distribution around its mean $\eta$ position), is then
exploited. Direct photons have a smaller cluster width, while background photons tend to have wider widths. A binned
maximum likelihood fit of the data to two contributions is performed: one accounting for the signal photons (modelled with Pythia~\cite{Pythia} embedded in PbPb data) and another one for the background, taking a template from the PbPb data. The number of isolated prompt photons is extracted from the fit.

\begin{figure}[hbtp]
  \begin{center}
    \includegraphics[width=0.45\linewidth]{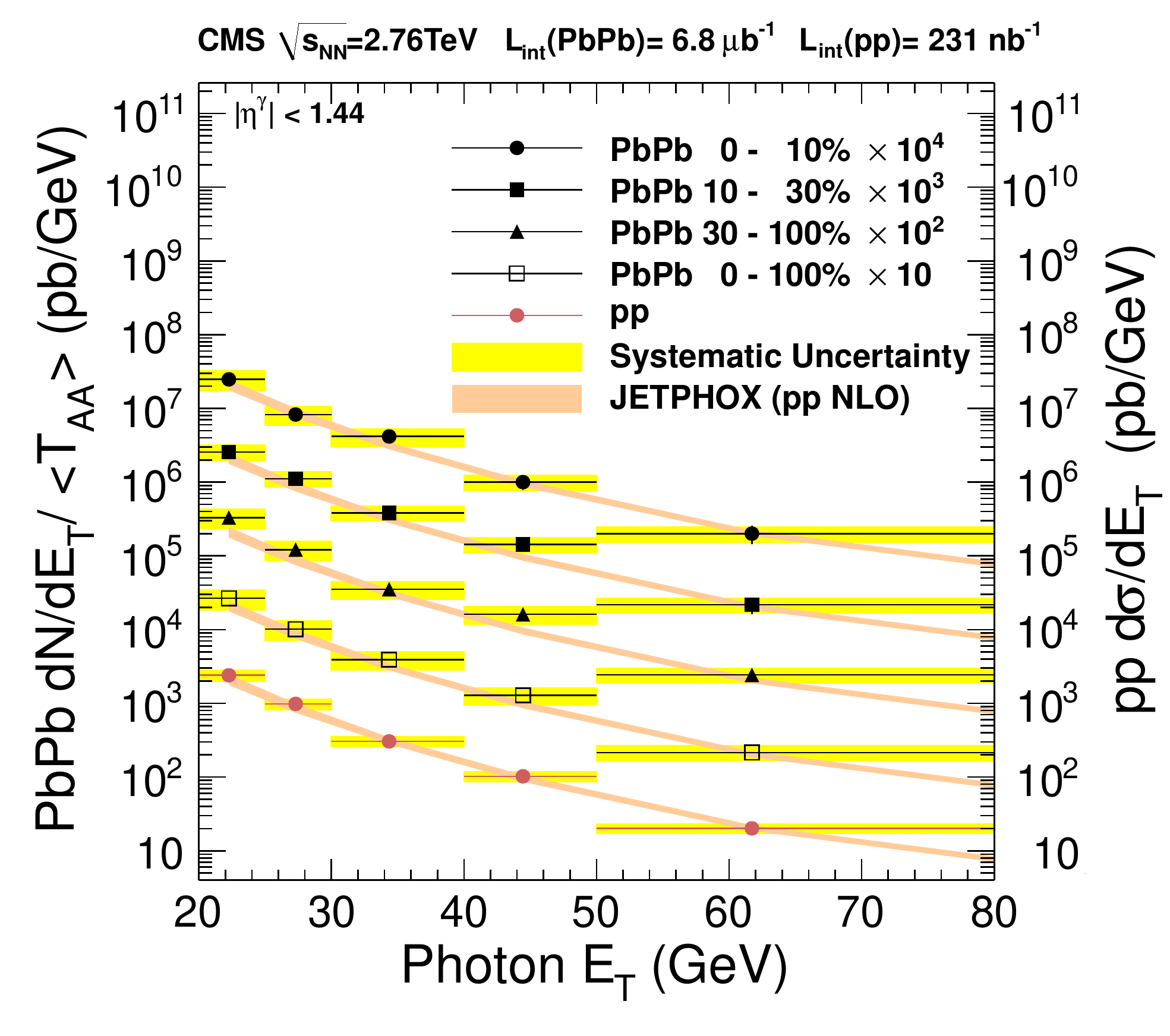} \includegraphics[width=0.45\linewidth]{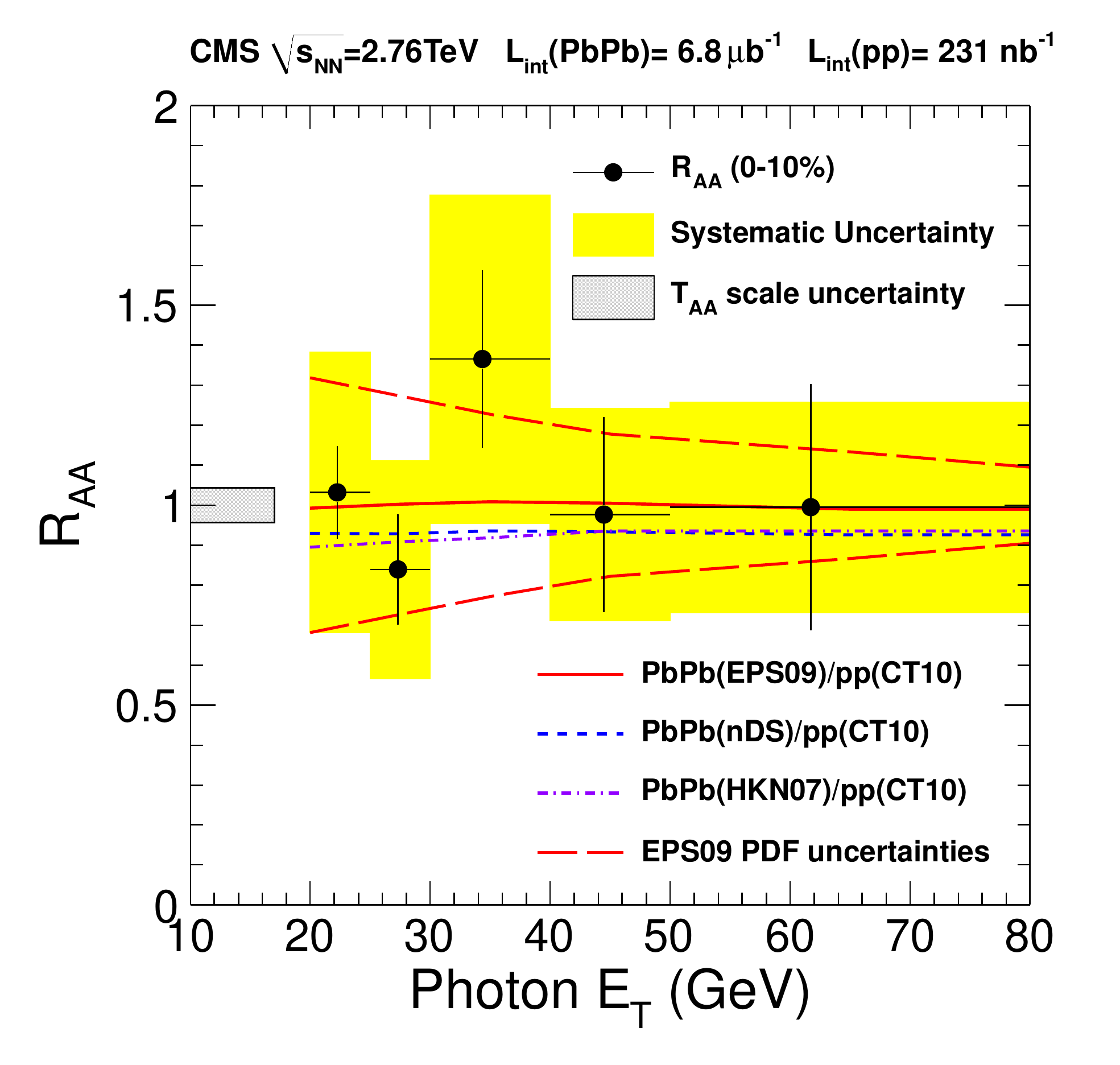}% 0.95 for PRL
    \caption{Left: Isolated photon spectra as a function of the photon E$_T$ for several centrality regions in PbPb collisions (scaled by T$_{\rm {AA}}$) and pp collisions, scaled by arbitrary factors (for better visualization). The results are compared to the NLO JETPHOX calculations. Right: Nuclear modification factors, R$_{AA}$, as a function of the photon E$_{\rm T}$ for the 0-10\% most central PbPb collisions at $\sqrt{s}= 2.76$ TeV .   }
    \label{fig:isol_phot_energy}
  \end{center}
\end{figure}

Figure~\ref{fig:isol_phot_energy} (left) shows the isolated photon yield per unit of photon transverse energy, E$_{\rm T}$, derived
from PbPb data and normalised to the number of binary collisions (N$_{\rm {coll}}$) by means of the nuclear factor T$_{AA}$\footnote{T$_{AA}$ is the number of elementary binary nucleon-nucleon (NN) collisions, N$_{\rm {coll}}$, divided by the elementary NN cross section.}.
This quantity is presented as a function of the photon E$_{\rm T}$ and for three different regions in centrality ( [0--10\%],
[10--30\%] and [30--100\%]) as well as the whole minimum bias sample. The cross sections are extracted for the photon
pseudorapidity region $|\eta^{\gamma}|< 1.44$, the ECAL barrel region. The corresponding cross section obtained
in pp data, at the same centre-of-mass energy, is also included. The photon energy distributions from PbPb data, scaled by T$_{AA}$, agree
 quite well with those from pp collisions and with NLO JETPHOX~\cite{JETPHOX} calculations.

 From this comparison, nuclear modification factors, R$_{\rm AA} = \sigma_{\rm PbPb}^{\rm norm} (\gamma) / \sigma_{\rm pp} (\gamma)$ are derived, where $\sigma_{\rm PbPb}^{\rm norm} (\gamma)$ indicates the photon yield in PbPb data, divided by the normalising factor T$_{\rm AA}$. This is done for different centrality regions, as a function of photon E$_{\rm T}$. Figure~\ref{fig:isol_phot_energy} (right) shows the [0-10 \%] case, where the data points are compared to several model predictions for the nuclear PDF
(EPS09~\cite{salgado_phot}, nDS~\cite{nDS} and HKN07~\cite{HKN07}).
The error bars account for the statistical uncertainty
and the coloured-filled rectangles for the systematic ones, which come mainly from the background template modelling and the photon energy scale, both in PbPb and in pp data. The R$_{\rm AA}$ values are compatible with unity and, thus, confirm the scaling of the isolated photon production with
 N$_{\rm {coll}}$. The present uncertainties preclude any quantitative constraint on the (small)
 nuclear effects predicted by the different theoretical models.

\begin{figure}[hbtp]
  \begin{center}
    \includegraphics[width=0.62\linewidth]{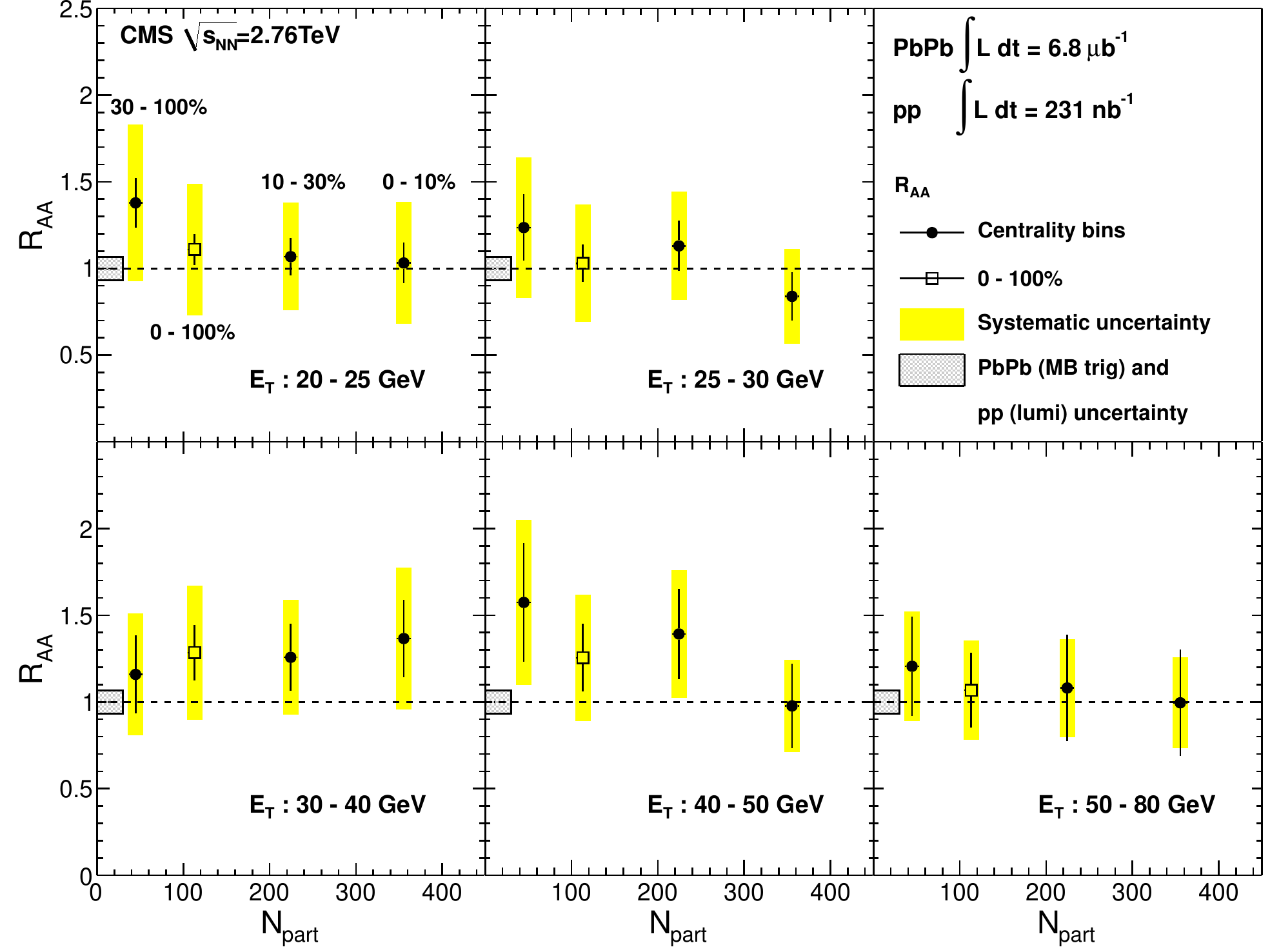} % 0.95 for PRL
    \caption{Nuclear modification factors, R$_{AA}$, as a function of PbPb centrality (given by the number of participant nucleons, N$_{\rm {part}}$) for five different photon E$_{\rm T}$ intervals. }
    \label{fig:isol_phot_allRAA}
  \end{center}
\end{figure}

The dependence of the R$_{AA}$ factors with the number of participants in the collision, N$_{\rm {part}}$, for different photon E$_{\rm T}$ bins, can be seen in Figure~\ref{fig:isol_phot_allRAA}.

These results allow for the establishment of the isolated photon sample as a baseline for further analyses, like for example, the production of photons accompanied by jets or, with larger data samples, constrain the nuclear PDF.

\section{Z bosons}
\label{Zboson}

\begin{figure}[hbtp]
  \begin{center}
    \includegraphics[width=0.627\linewidth]{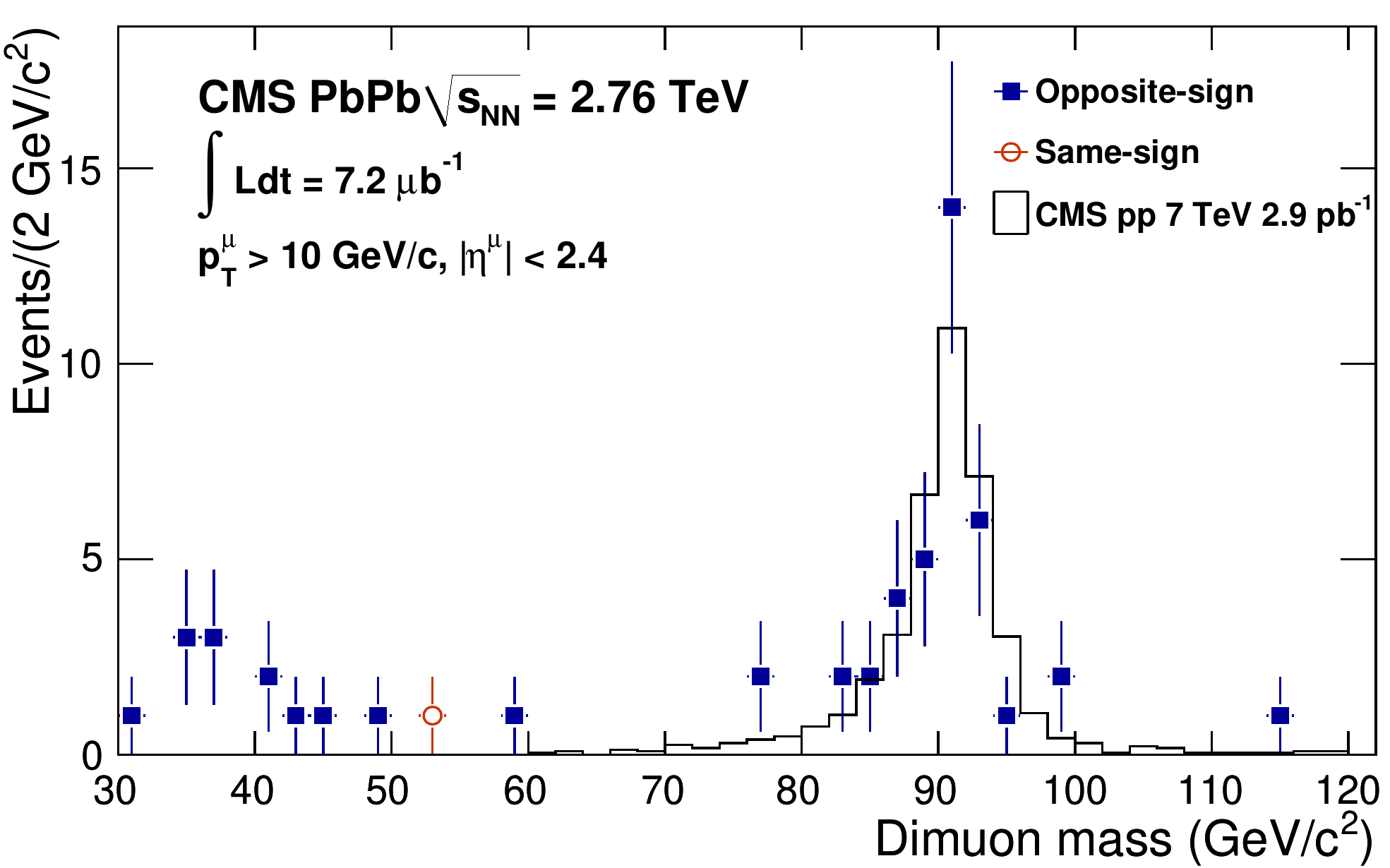} % 0.95 for PRL
    \caption{  Dimuon invariant mass spectra in PbPb collisions. Full squares are opposite-sign dimuons, while the empty circle shows a unique like-sign dimuon candidate. The histogram shows the corresponding distribution measured in pp collisions at 7 TeV within 60-120 GeV/c$^2$, scaled to the 39 PbPb candidates.}
    \label{fig:Z_dimuon_mass}
  \end{center}
\end{figure}

The Z bosons have been detected in CMS in their Z$\rightarrow \mu^+\mu^-$ and Z$\rightarrow e^+e^-$ decay channels.
In the following, results in the muonic decay channel, based on an integrated luminosity ${\cal L} = 7.2\, {\rm pb}^{-1}$ of PbPb collisions data, are presented. The selection in this channel exploits their quite clear signature: two oppositely charged high transverse momenta muons (p$_{\rm T}^{\mu} > 10$ GeV/c). In this way, 39 Z candidates are extracted, with an almost negligible background. Figure~\ref{fig:Z_dimuon_mass} presents the dimuon invariant mass. The Z mass peak in PbPb data has a resolution similar to that obtained in pp data.

\begin{figure}[hbtp]
  \begin{center}
    \includegraphics[width=0.45\linewidth]{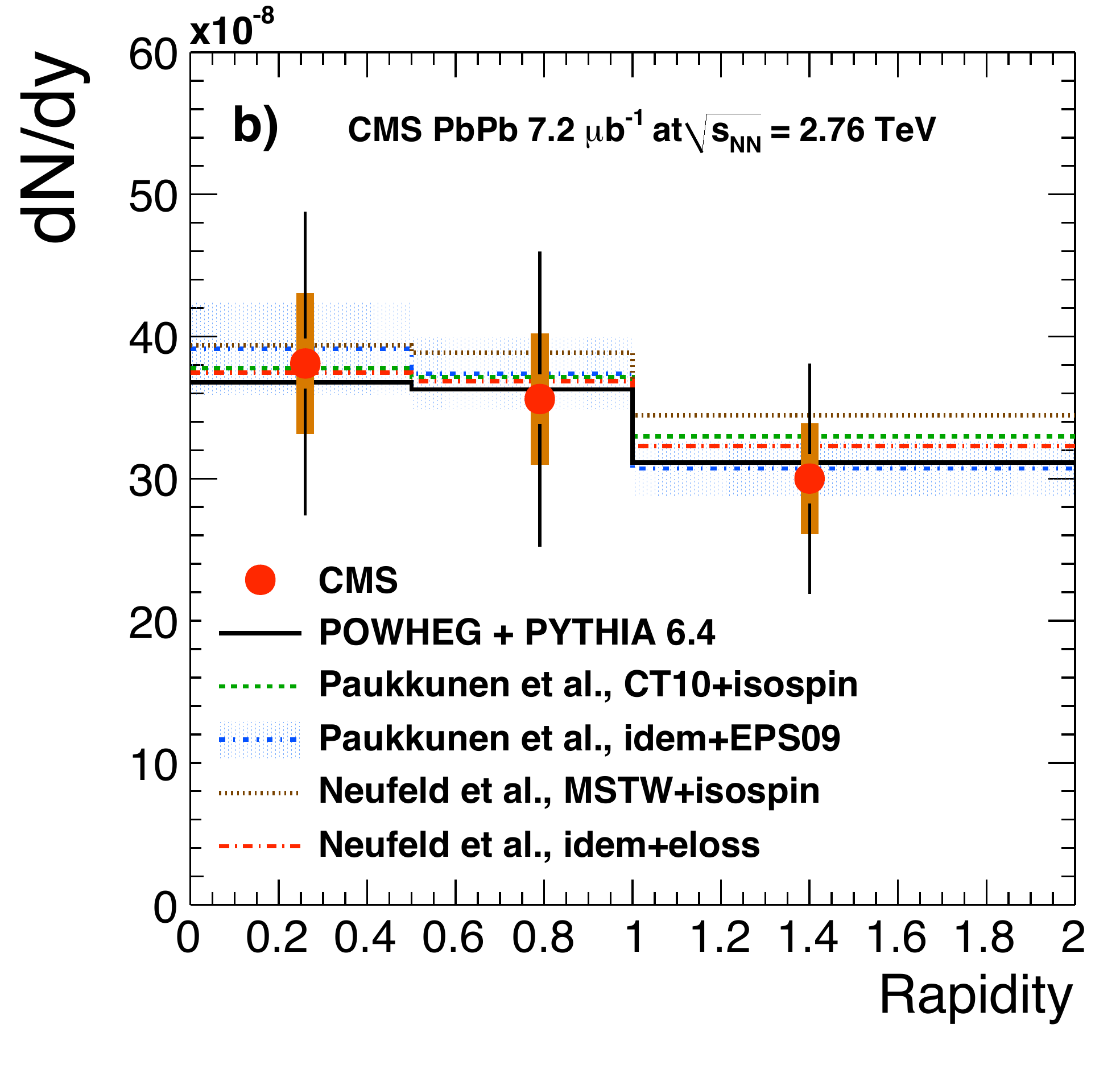}  \includegraphics[width=0.45\linewidth]{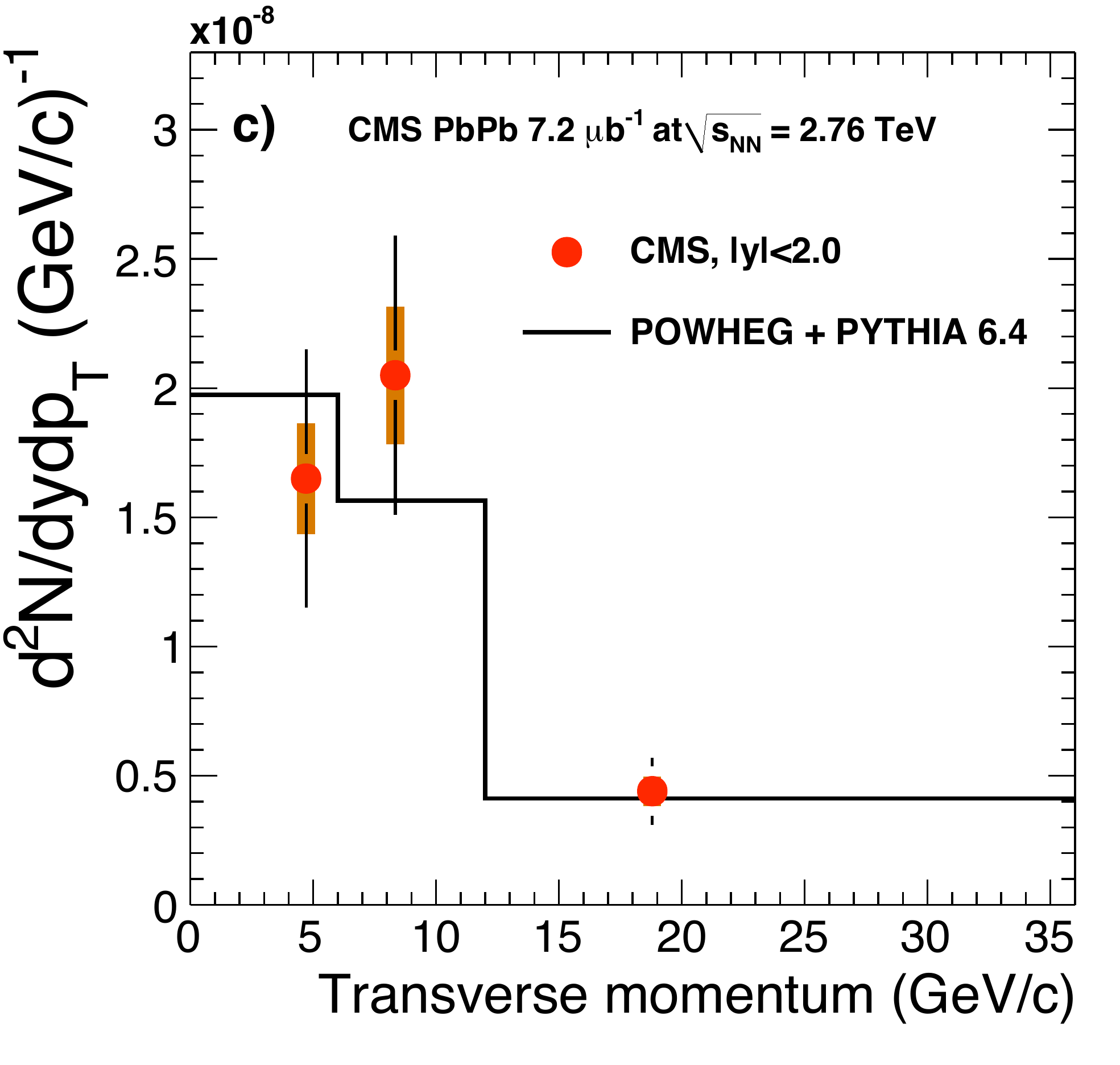}
    \caption{Yields of Z $\rightarrow \mu \mu$ per event in PbPb collisions: Left: dN/dy versus Z boson rapidity, y. Right: d$^2$N/dydp$_{\rm T}$ versus the Z boson p$_{\rm T}$. Vertical lines (bands) correspond to statistical (systematic) uncertainties. See References section for the different nuclear PDF models. }
    \label{fig:Z_rap_pt}
  \end{center}
\end{figure}

Several kinematic distributions have been studied with this sample of Z candidates, like the Z rapidity and p$_{\rm T}$, shown in Figure~\ref{fig:Z_rap_pt} left and right, respectively. They are compared to Powheg~\cite{Powheg} predictions (perturbative QCD at NLO) for pp $\rightarrow {\rm Z} \rightarrow \mu^+\mu^-$, scaled with the number of binary collisions, and also to other models simulating PbPb interactions which incorporate nuclear effects such as isospin, energy loss and shadowing\cite{Salgado_Z, Neufeld}
 The experimental distributions are consistent with them all, not being still precise enough to quantitaively distinguish between them.

\begin{figure}[hbtp]
  \begin{center}
    \includegraphics[width=0.5\linewidth]{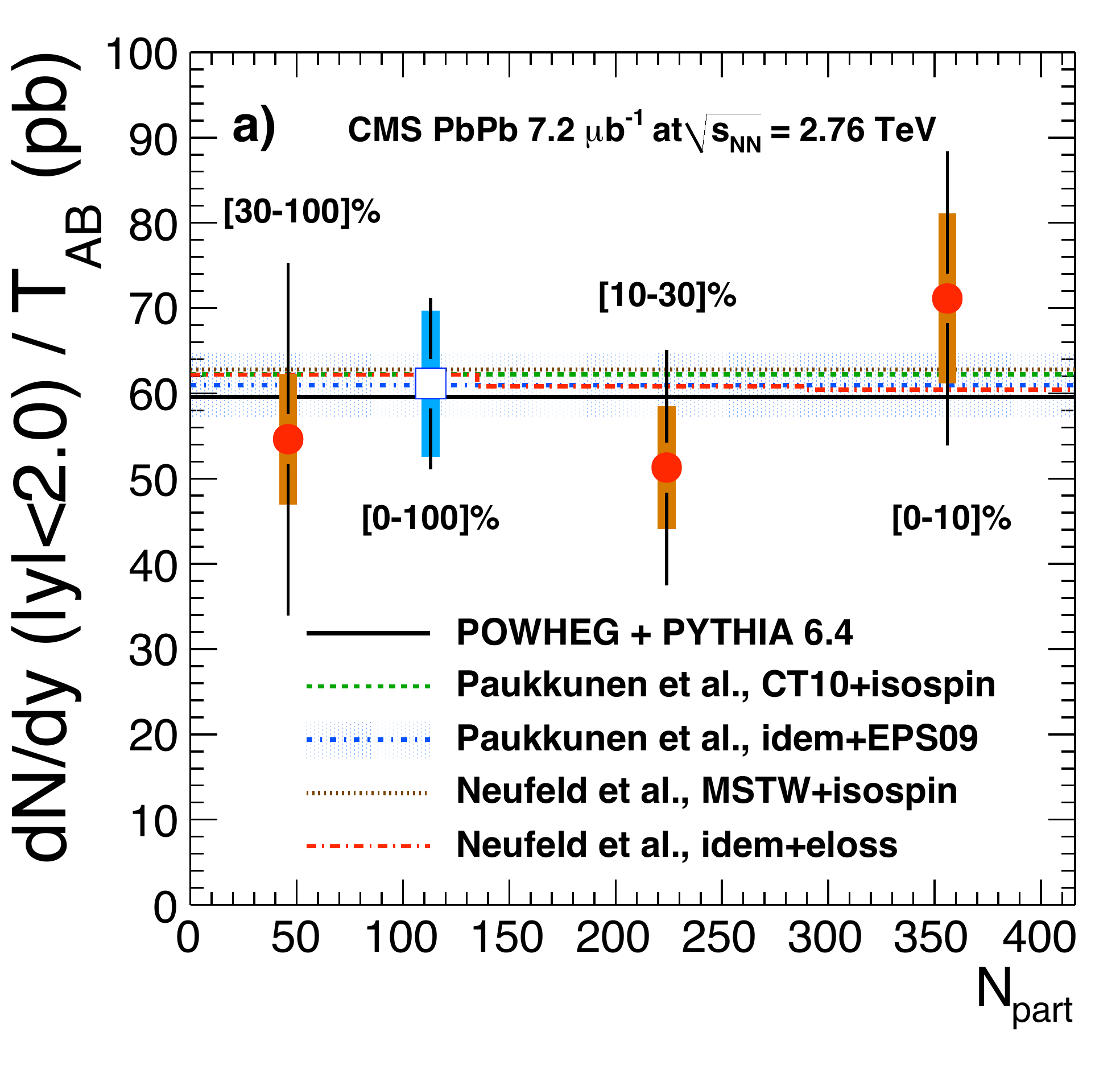} % 0.95 for PRL
    \caption{Centrality dependence (as a function of N${\rm {part}}$) of the yield of Z $\rightarrow \mu \mu$ per event divided by the expected nuclear overlap function T$_{\rm AA}$. Vertical lines (bands) correspond to statistical (systematic) uncertainties. See References section for the different nuclear PDF models. }
    \label{fig:Z_Npart}
  \end{center}
\end{figure}

When attending to the centrality of the PbPb collision, no dependence of the production of Z bosons on the number of participants is observed, as shown in Figure~\ref{fig:Z_Npart}. Using the prediction for pp $\rightarrow {\rm Z} \rightarrow \mu^+\mu^-$ from Powheg, scaled with the number of binary collisions, and the minimum bias yield of Z $\rightarrow \mu^+\mu^-$ in PbPb interactions, a relation R$_{\rm AA} = 1.00 \pm 0.16 \pm 0.14 $ is established, validating the Z bosons as a reference particle unaffected by final-state interactions in the strongly dense system produced in Pb-Pb
collisions at the LHC.
.

\section{W bosons}
\label{Wboson}

 W bosons have been detected and measured at the LHC for the first time in heavy ion collisions, in their W$^\pm \rightarrow \mu^\pm \nu$ decay channel. Their distinctive experimental feature is the presence of a single high p$_{\rm T}$ muon recoiling in the transverse plane to a significant imbalance in the total momentum of the event, which accounts for the undetected neutrino.

\begin{figure}[hbtp]
  \begin{center}
    \includegraphics[width=0.46\linewidth]{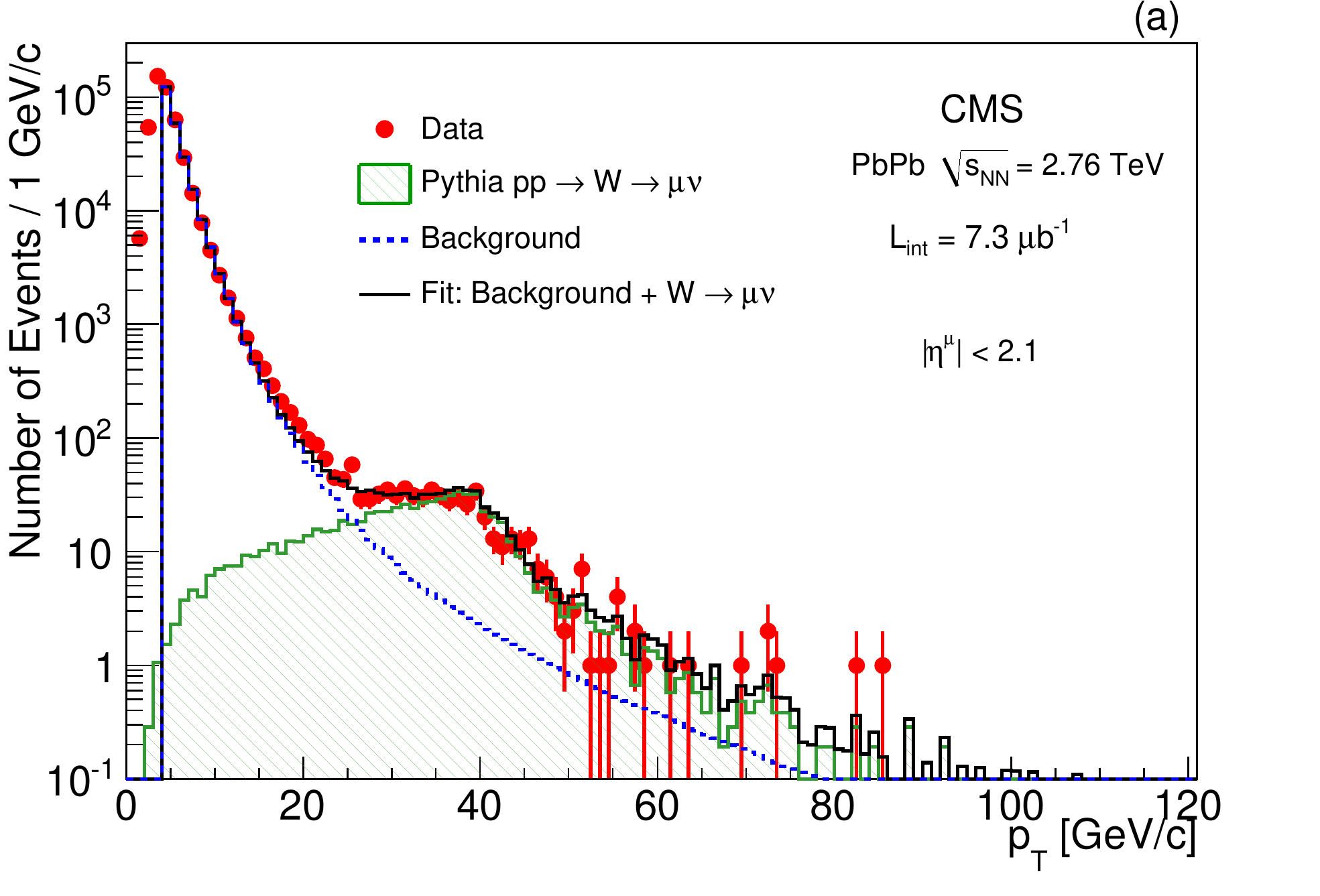} \includegraphics[width=0.46\linewidth]{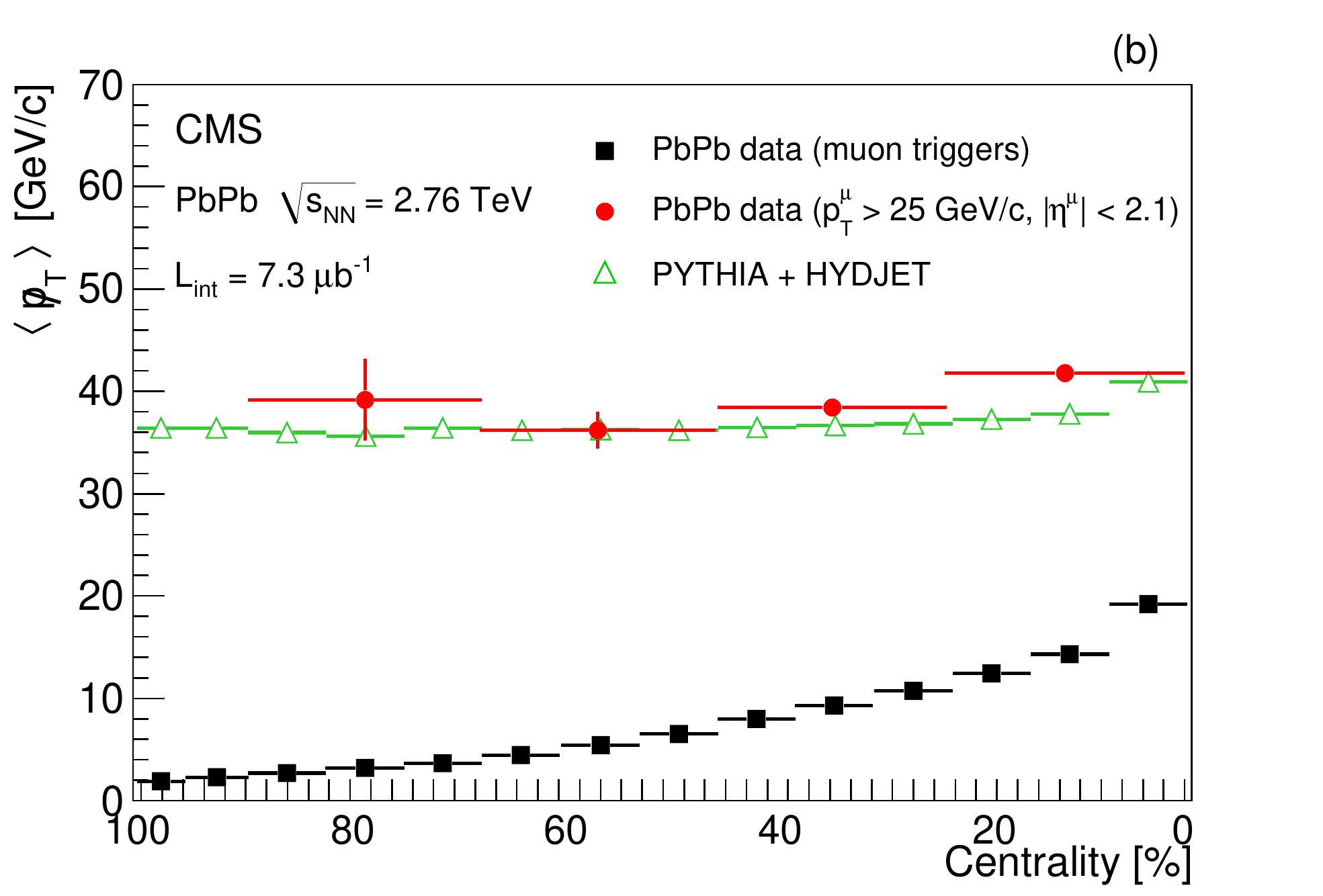} % 0.95 for PRL
    \caption{Left: Single-muon transverse-momentum spectrum for $|\eta^{\mu}| < 2.1$ in PbPb data (red points). Signal (green-hatched histogram) and background (blue-dashed histogram) contributions are fitted (black-solid line) to the data. Right: Mean value of $p\!\!\!/_{\rm \!T}$ for charged tracks as a function of the event centrality, before any selection on muon-triggered data (black squares) and after it (red circles), together with predictions from PYTHIA + HYDJET~\cite{HYDJET} (green triangles). }
    \label{fig:W_pt_muon}
  \end{center}
\end{figure}

Figure~\ref{fig:W_pt_muon} (left) shows the muon transverse momentum distribution (solid dots) for events selected online with a low (2-3 GeV/c) muon p$_{\rm T}^{\mu}$ threshold and fulfilling some quality cuts, such as a minimum number of hits used in their offline reconstruction or requiring a distance in the transverse plane between the muon impact parameter and the primary vertex to be less than 300 $\mu m$. The muon pseudorapidity region studied is $|\eta^\mu|< 2.1$. A Z-veto is applied based on the presence of a second high-p$_{\rm T}$ muon and the resulting dimuon invariant mass. An enhancement of events is observed in the region p$_{\rm T}^{\mu} > 25$ GeV/c, where the contribution from muons from W and Z decays is expected. In fact, a fit (black solid line) to the data is performed, taking into account two contributions, one for the W$\rightarrow \mu \nu$ signal (modelled with a Pythia pp $\rightarrow {\rm W} \rightarrow \mu \nu$ simulation, the normalization left as a free parameter in the fit) (green-hatched histogram) and another for the background, described by a 3 parameter function (blue-dashed line). The fit describes the data quite well.

The imbalance in the transverse momentum in the event, ($p\!\!\!/_{\rm \!T}$), is computed as the opposite sign of the vectorial sum of all charged particle transverse momenta in the event, with p$_{\rm T} > 3$ GeV/c. The mean value of this quantity for all online selected muon events is shown in Figure~\ref{fig:W_pt_muon} (right) (black squares) as a function of the event centrality. When requiring the presence of a high-p$_{\rm T}$ muon (p$_{\rm T}^{\mu} > 25$ GeV/c) the associated mean value of $p\!\!\!/_{\rm \!T}$ shifts to higher values of  $ \approx 40$~GeV/$c$, and is far less dependent on the centrality of the collision. This result agrees with that expected for the typical $p\!\!\!/_{\rm \!T}$ produced by an undetected neutrino originating from W decay. To enhance the contribution from the W signal, events are therefore required to have $p_{\rm T}^\mu > 25$~GeV/$c$ and $p\!\!\!/_{\rm \!T} > 20$~GeV/$c$.

\begin{figure}[hbtp]
  \begin{center}
    \includegraphics[width=0.61\linewidth]{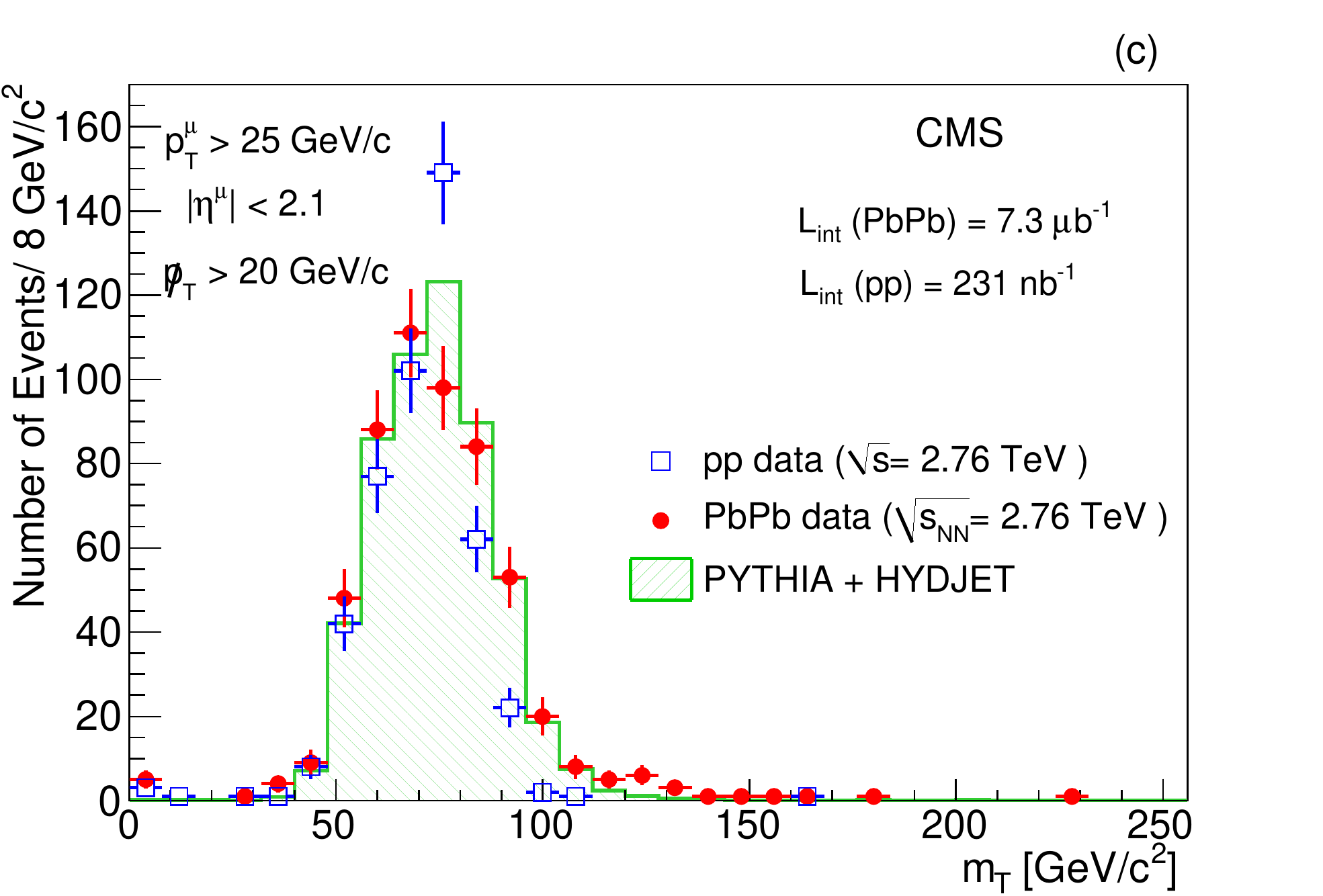}  % 0.95 for PRL
    \caption{Transverse mass distribution for selected events in PbPb (red-filled circles) and pp (blue open squares) data, compared to PYTHIA + HYDJET simulation (hatched histogram). The error bars represent statistical uncertainties.}
    \label{fig:W_mt}
  \end{center}
\end{figure}

With this selection, the transverse mass distribution, given by
$m_{\rm T} = \sqrt{2 p_{\rm T}^\mu p\!\!\!/_{\rm \!T} ( 1 - \cos \phi)}$, being $\phi$ the difference in azimuthal angle between the muon and $p\!\!\!/_{\rm \!T} $ vectors, of the remaining events is presented in Figure~\ref{fig:W_mt} for PbPb (red dots). The data sample of pp collisions collected at the same centre-of-mass energy is analyzed in a similar way, and their $m_{\rm T}$ distribution is also shown in the figure (blue-open squares). They both represent compatible W signals, almost background free, as they are in good agreement with a W Pythia signal simulation, embedded in PbPb events generated with Hydjet (PYTHIA + HYDJET~\cite{HYDJET}) (green-hatched histogram).
Residual contamination from other electroweak processes (Z$\rightarrow \mu^+\mu^-$ and W$^\pm \rightarrow \tau^\pm \nu$) (2\%) are subtracted, and QCD background is estimated to be $\approx 1\%$ and included as a systematic uncertainty.

\begin{figure}[hbtp]
  \begin{center}
    \includegraphics[width=0.46\linewidth]{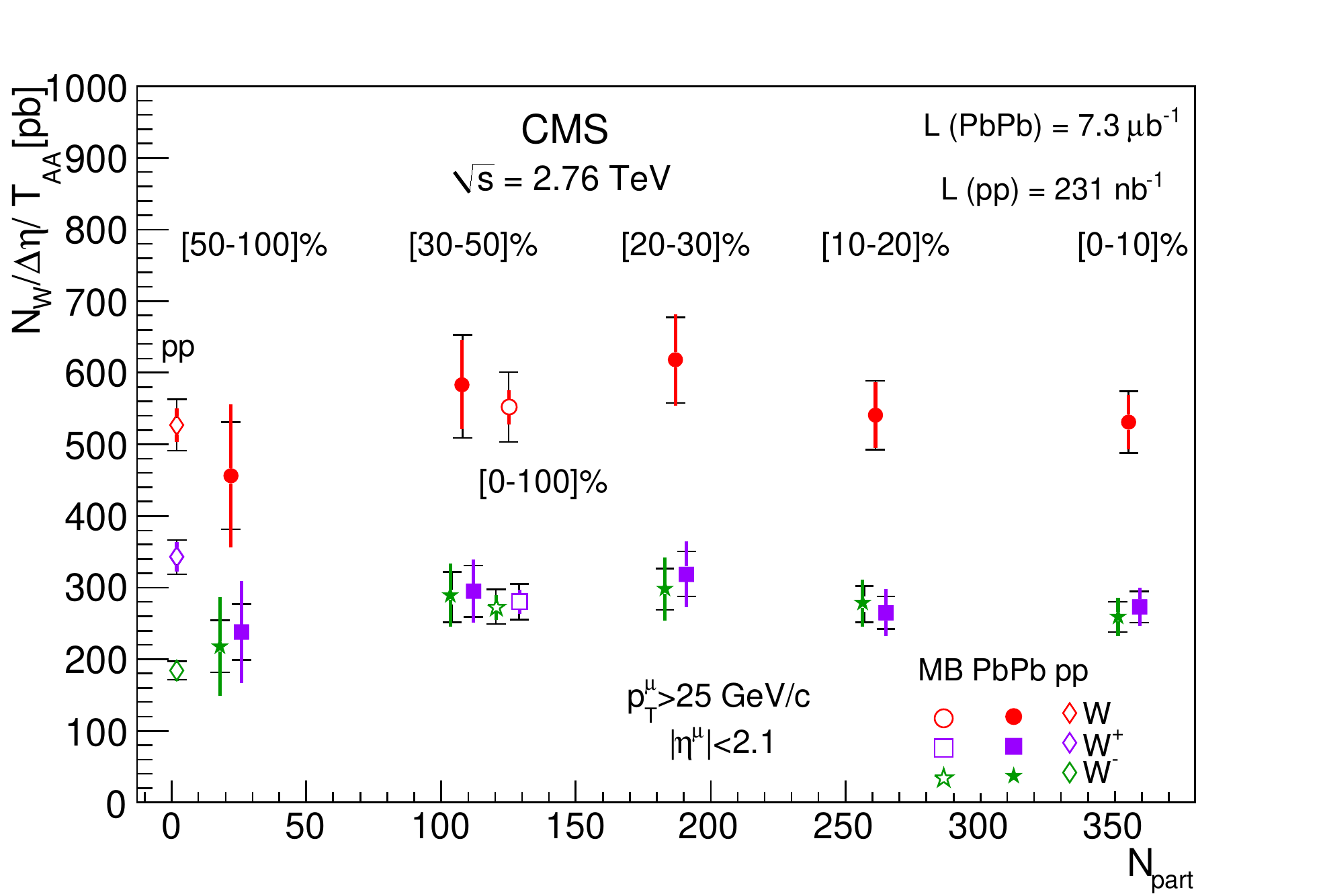} \includegraphics[width=0.46\linewidth]{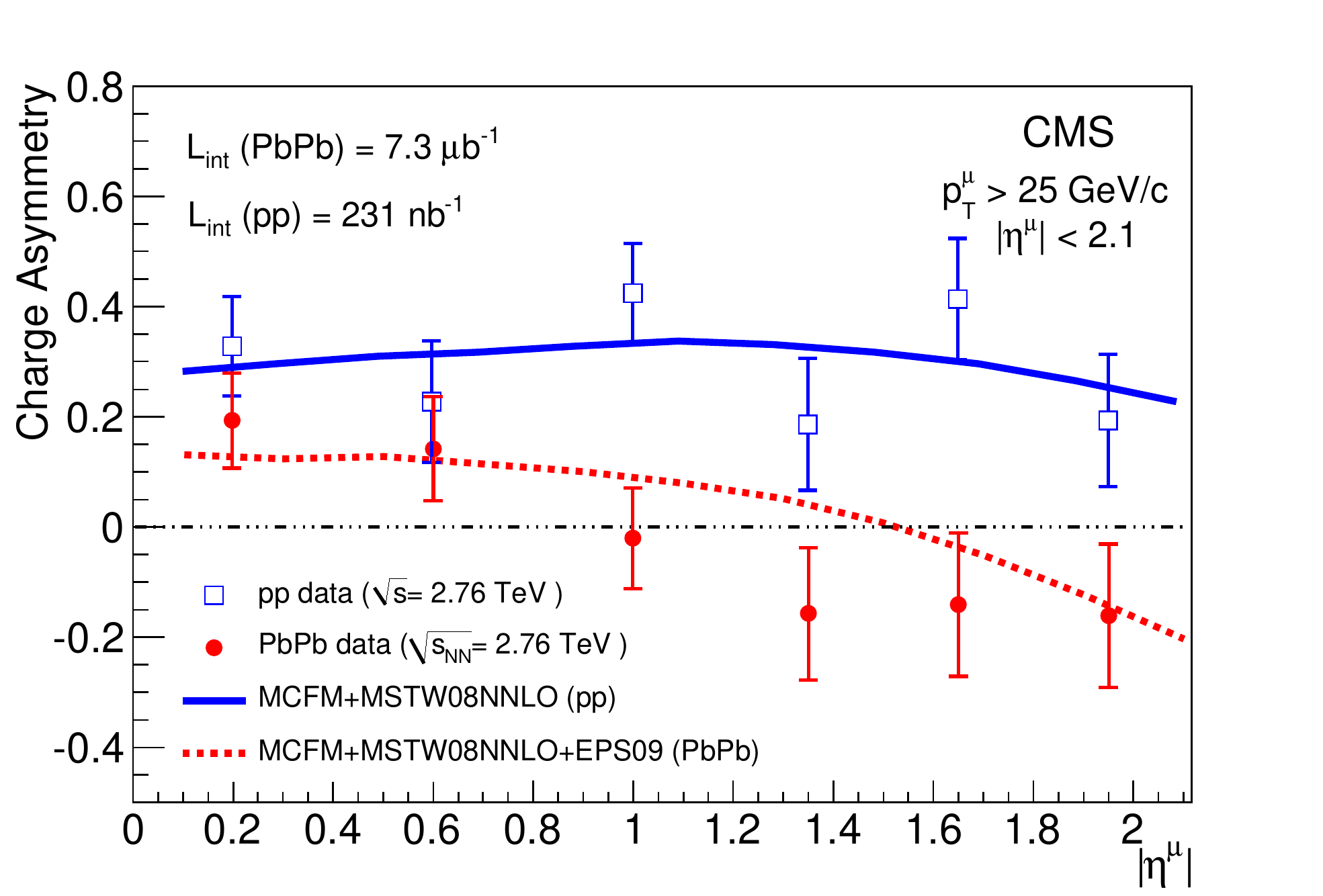} % 0.95 for PRL
    \caption{Left: Centrality dependence of normalised W $\rightarrow \mu\nu $ cross sections in PbPb collisions, for all W candidates (red filled points) and separated by charge, W$^+$ (violet-filled squares) and W$^-$ (green-filled stars). The open symbols at N$_{\rm part} = 120$ represent the minimum bias events. At N$_{\rm part} = 2$, the corresponding cross sections are displayed for pp collisions for the same $\sqrt{s}$. The cross sections are given for the phase space region $p_{\rm T}^\mu > 25$~GeV/$c$ and $|\eta^\mu|< 2.1$. Right: Charge asymmetry as a function of muon $|\eta|$ for PbPb (red-filled circles) and pp (open blue squares) collisions. Overlaid are predictions from MCFM calculations with MSTW08 PDF (blue curve) and additionally with EPS09 PDF for PbPb collisions. }
    \label{fig:W_cros_section}
  \end{center}
\end{figure}

From the sample of W candidates selected, 275(301) W$^+$ and 264(165) W$^+$ in PbPb(pp) data, the W production cross section in PbPb collisions, scaled by the nuclear factor T$_{\rm AA}$ and in pp interactions, per unit of W rapidity, is computed and shown in Figure~\ref{fig:W_cros_section} (left) as a function of the number of participants in the collision, N$_{\rm {part}}$. Two main effects are observed: one, the W production in PbPb does not depend on the centrality of the collision and the other, the comparison of the individual W$^+$ and W$^-$ production in PbPb and pp data, reflect the different u and d quark content in Pb nuclei and in the proton. The R$_{\rm AA}$ factors obtained are: R$_{\rm AA}(\rm W^+) = 0.82 \pm 0.07 \pm 0.09$ and R$_{\rm AA}(\rm W^-) = 1.46 \pm 0.14 \pm 0.16$. Nevertheless, the total W (W$^+$ plus W$^-$) production is consistent, at LO, with N$_{\rm {coll}}$ scaling, as R$_{\rm AA}(\rm W) = 1.04 \pm 0.07 \pm 0.12$

The difference in W$^+$ (u$\bar{\rm d} \rightarrow {\rm W}^+$) and W$^-$ (d$\bar{\rm u} \rightarrow {\rm W}^-$) production at LHC and the spin conservation rule in their leptonic decay originate a different angular yield of W$^+ \rightarrow \mu^+ \nu$ and W$^- \rightarrow \mu^- \nu$, known as charge asymmetry and given by
$A = (N_{W^+} - N_{W^-}) / (N_{W^+} + N_{W^-})$, being N$_{W}$ the efficiency corrected yield of W candidates. The distribution of the asymmetry as a function of the muon $|\eta|$ is presented in Figure~\ref{fig:W_cros_section} (right) for both PbPb and pp collisions at $\sqrt{s} = 2.76$ TeV, together with predictions from MCFM~\cite{MCFM} generator using MSTW08~\cite{MSTW08} PDF and in addition, for the PbPb data, the EPS09~\cite{Salgado_Z} nuclear PDF.

\begin{figure}[hbtp]
  \begin{center}
    \includegraphics[angle=90,width=0.53\linewidth]{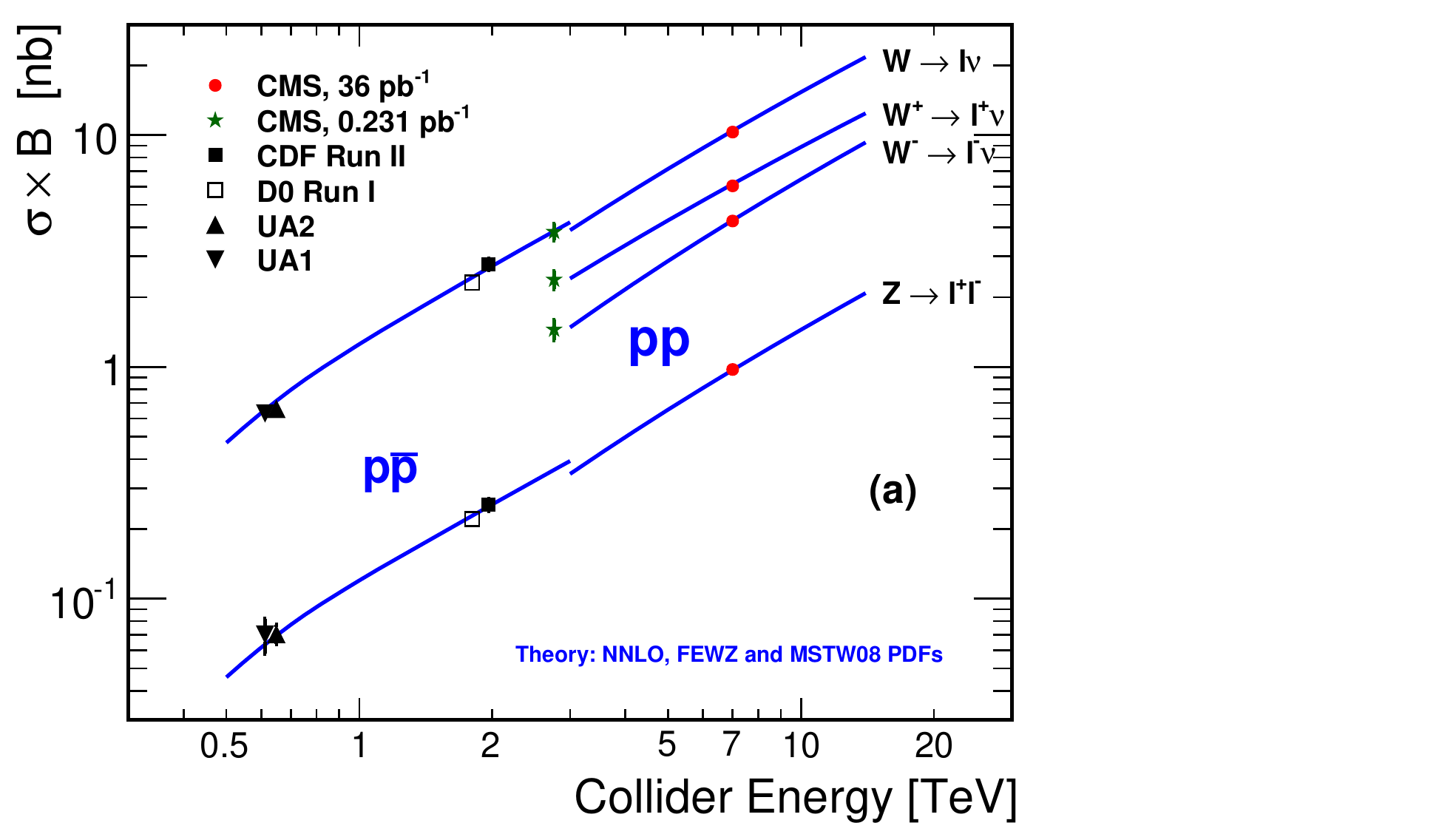} \includegraphics[width=0.38\linewidth]{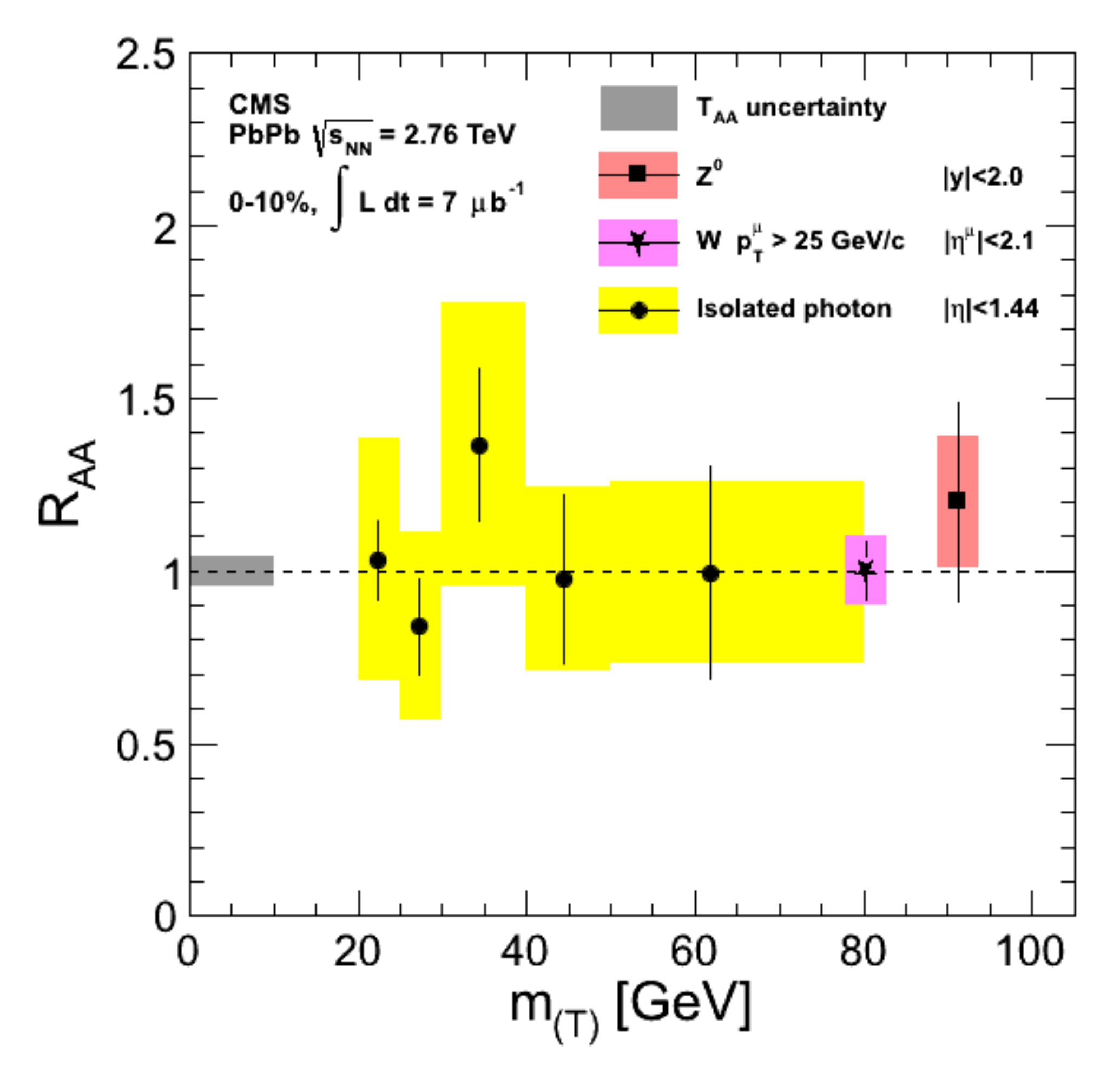} % 0.95 for PRL
    \caption{Left: W (and W$^+$, W$-$)  production cross section as a function of $\sqrt{s}$ for pp / p$\bar{\rm p}$ collisions, as measured by different experiments. Right: Nuclear modification factor, R$_{\rm AA}$, for W and Z bosons, and for the isolated photons sample, in bins of transverse energy/mass.}
    \label{fig:all_colliders}
  \end{center}
\end{figure}

As a conclusion a new measurement has been provided of the W production cross sections at a different centre-of-mass energy in the pp colliding system, validating once more the Standard Model prediction, as observed in Figure~\ref{fig:all_colliders} (left) where the W production cross section is depicted as a function of $\sqrt{s}$. For the first time, the W yield has been measured in PbPb interactions, establishing the W as a reference particle unaffected by the hot and dense medium created, as seen in  Figure~\ref{fig:all_colliders} (right), where the R$_{\rm AA}$ nuclear factor for the W as for the rest of electroweak bosons (Z and isolated photons) are shown to be consistent with unity over a wide range of energies, establishing these particles as "clean" probes of the initial state of the collision.

Whenever the uncertainties (both statistical and systematic) will allow it, the analysis of these bosons will be very useful to constrain predicted nuclear effects (shadowing, energy loss) and give access to quark/antiquark/gluon parton density functions (PDF) in the proton and neutron.

\providecommand{\href}[2]{#2}\begingroup\raggedright\begin{thebibliography}{}%
\makeatletter
\providecommand{\hrefCMSnoop }[0]{\@secondoftwo}%
\makeatother
\providecommand{\doi}{\texttt{doi:}\begingroup \urlstyle{tt}\Url}

\end{thebibliography}\endgroup


\begin{thebibliography}{00}


\bibitem{Adler:2005ig}
      PHENIX Collaboration, {\em Phys. Rev. Lett.} {\bf 94} (2005) 232301, {\tt doi:10.1103/PhysRevLett.94.232301}, {\tt arXiv:nucl-ex/0503003}.

\bibitem{Chatrchyan:2011ip}
      CMS Collaboration, {\em Phys. Lett. B.} {\bf 716} (2012) 256, {\tt doi:10.1016/j.physletb.2012.02.077}, {\tt arXiv:1201.3093}.

\bibitem{CMS_gamma_jet}
      CMS Collaboration, Submitted to {\em Phys. Lett. B.}, {\tt arXiv:1205.0206}.

\bibitem{Atlas:2010px}
      ATLAS Collaboration, {\em Phys. Lett. B.} {\bf 697} (2011) 294, {\tt doi:10.1016/j.physletb.2011.02.006}, {\tt arXiv:1012.5419}.

\bibitem{Chatrchyan:2011ua}
      CMS Collaboration, {\em Phys. Rev. Lett.} {\bf 106} (2011) 212301, {\tt doi:10.1103/PhysRevLett.106.212301}, {\tt arXiv:1102.5435}.

\bibitem{W-paper}
      CMS Collaboration, {\em Phys. Lett. B.} {\bf 715} (2012) 66, {\tt doi:10.1016/j.physletb.2012.07.025}. 

\bibitem{CMS_det}
      CMS Collaboration, {\em JINST} {\bf 03} (2008) S08004, {\tt doi:10.1088/1748-0221/3/08/S08004}.

\bibitem{Pythia}
      Sj{\"o}strand, T. and Mrenna, S. and P. Skands, {\em JHEP} {\bf 05} (2006) 026, {\tt doi:10.1088/1126-6708/2006/05/026}, {\tt arXiv:hep-ph/0603175}.

\bibitem{JETPHOX}
      S. Catani, M. Fontannaz, J. Guillet, {\em JHEP} 0205 (2002) 028, {\tt doi:10.1088/1126-6708/2002/05/028}.

\bibitem{salgado_phot}
      K.J. Eskola, H. Paukkunen, C.A. Salgado, {\em  JHEP} 0904 (2009) 065 {\tt doi:10.1088/1126-6708/2009/04/065}, {\tt arXiv:0902.4154}.

\bibitem{nDS}
     D. de Florian, R. Sassot, {\em Phys. Rev. D} {\bf 69} (2004) 074028, {\tt  doi:10.1103/PhysRevD.69.074028}, {\tt arXiv:hep-ph/0311227}.

\bibitem{HKN07}
     M. Hirai, S. Kumano, T.-H. Nagai, {\em Phys. Rev. C} {\bf 76} (2007) 065207, {\tt doi:10.1103/PhysRevC.76.065207}, {\tt arXiv:0709.3038}.

\bibitem{Powheg}
     S. Alioli, P. Nason, C. Oleari, and E. Re, {\em JHEP} 07 (2008) 060, {\tt doi:10.1088/1126-6708/2008/07/060}, {\tt arXiv:0805.4802}.

\bibitem{Salgado_Z}
     H. Paukkunen and C. A. Salgado, {\em JHEP} 03 (2011) 071, {\tt doi:10.1007/JHEP03(2011)071}, {\tt arXiv:1010.5392}.

\bibitem{Neufeld}
     R. B. Neufeld, I. Vitev, and B.W. Zhang, {\em Phys. Rev. C} {\bf 83} (2011) 034902, {\tt doi:10.1103/PhysRevC.83.034902}, {\tt arXiv:1006.2389}.

\bibitem{HYDJET}
    Lokhtin, I. P. and Snigirev, A. M., {\em Eur. Phys. J. C} {\bf 45} (2006) 211, {\tt doi:10.1140/epjc/s2005-02426-3}, {\tt arXiv:hep-ph/0506189}.

\bibitem{MCFM}
    J. Campbell, K. Ellis and C. Williams, {\em JHEP} 07 (2011) 018, {\tt doi:10.1007/JHEP07(2011)018}, {\tt arXiv:1105.0020}.

\bibitem{MSTW08}
    Martin, A. D. and Stirling, W. J. and Thorne, R. S. and Watt, G., {\em Eur. Phys. J. C} {\bf 63} (2009) 189, {\tt doi:10.1140/epjc/s10052-009-1072-5}, {\tt arXiv:0901.0002}.

\end{thebibliography}
\end{document}